\documentstyle[twoside,fleqn,espcrc2,epsf]{article}

\newcommand{\beq}{\begin{equation}}
\newcommand{\eeq}{\end{equation}}
\newcommand{\bea}{\begin{eqnarray}}
\newcommand{\eea}{\end{eqnarray}}
\newcommand{\ol}[1]{\overline{#1}}

\newcommand{\ONE}{\makebox[0pt][l]{\hspace{0.05cm}1}1}
\newcommand{\tr}{{\rm Tr}}

\title{ 
\vspace{-3.5cm} 
\begin{flushright}
{\normalsize\tt OHSTPY-HEP-T-01-028}\\
\end{flushright}
\vspace*{2.0cm}
Heavy-light physics using NRQCD-staggered actions
} 

\author{
Matthew Wingate\address{Department of Physics, Ohio State University,
	Columbus, Ohio 43210, USA}
\thanks{In collaboration with 
Junko Shigemitsu (Ohio State) and G.\ Peter Lepage
(Cornell). Talk given at {\it Lattice 2001}.
} 
}

\begin{document}

\begin{abstract}
One leading source of uncertainty in the lattice computation
of semi-leptonic form factors in $B$ decay,
and to a lesser extent $B$ meson decay constants,
comes from the extrapolation of the light quark mass to the
physical up/down mass.  This talk presents first explorations 
of simulating the light quark with staggered fermion actions, 
which are free of the spurious zero modes that effect Wilson-like
Dirac operators and are less costly.  Methods for fitting
to Euclidean-time NRQCD-staggered meson propagators are discussed,
and some preliminary spectrum results are presented. 
\end{abstract}

\maketitle

\section{INTRODUCTION}

Most of the research studying heavy-light mesons on the lattice
has focused on the formulation of heavy quarks; however systematic
effects due to the light quark are also important to investigate,
especially in decays such as $B\to \pi\ell\nu$.
So far Wilson-like actions have been used exclusively 
for the light quark action, and the appearance of 
``exceptional configurations'' can limit how small the light quark
mass can be set.  

This work reports on progress applying the familiar 
staggered action to the study of heavy-light
physics, using the usual NRQCD action for the heavy quark.
Although there are a few complications, such as 
opposite parity contributions to correlators, this method may prove
cheaper and more accurate than Wilson fermions
since staggered fermions have only 1 spin component and their cutoff
errors begin at $O(a^2)$ with 
straightforward tree-level improvement to $O(a^4)$.
Below we give a brief description of the method combining the
two actions, followed by preliminary spectrum results.  
Observations with improved actions are also made.

\section{METHOD}

It is well-known that the naive discretization 
of the QCD Lagrangian
\beq
{\cal L}_\Psi ~=~ \ol{\Psi}\big(\gamma\cdot\nabla + m \big) \Psi \, ,
\eeq
where
\bea
\nabla_\mu\, \Psi(x) &=& \frac{1}{2a} \, 
 [ U_\mu(x) \Psi(x+a\hat\mu) \nonumber \\
& - & U^\dagger_\mu(x - a\hat\mu) \Psi(x - a\hat\mu) ] \, ,
\eea
describes not 1 but 16 fermions due to the ``doubling'' symmetry.  
It is convenient to transform the fields
\beq
\Psi(x) \to \Omega(x)\Phi(x), ~~
\ol{\Psi}(x) \to \ol{\Phi}(x)\Omega^\dagger(x)
\eeq
with
$
\Omega(x) ~\equiv~ \prod_\mu ~(\gamma^\mu)^{\, x_\mu/a}
$
to obtain a spin-diagonal Lagrangian
\beq
{\cal L}_\Phi = \ol{\Phi}
\left[(\alpha\cdot\nabla + m)\times{\ONE} \,\right] \, \Phi 
\eeq
where
$
\alpha^\mu(x)\equiv (-1)^{(x_0+x_1+\ldots+x_{\mu-1})/a} \, .
$
Now the 16-fold degeneracy can be reduced to 4-fold
by defining fields $\chi$ with one spin-component:
\beq
\Phi(x) \equiv e(x) \chi(x) \, ;
\eeq
usually the c-number spinor is chosen to be constant
$e(x) = e$.  A simple identity between the $\Psi$ and
$\chi$ propagators can be derived:
\beq
G_\Psi(x;y) ~=~ G_\chi(x;y) \times 
\Omega(x)\Omega^\dagger(y) \, .
\label{eq:propident}
\eeq

Operators which couple to heavy-light meson states
can be constructed from a heavy field $Q(x)$
and the naive $\Psi(x)$.  $Q(x)$
may be described by a Wilson-like discretization or a
nonrelativistic formulation.
Considering local bilinear operators of the form:
\beq
J_{\Gamma}(x) ~ = ~ \ol{Q}(x)\Gamma\Psi(x)
\eeq
with
$\Gamma \in\{\ONE,\gamma^\mu,\gamma^\mu\gamma^\nu,\gamma^\mu\gamma^5,
\gamma^5\}$, we can see that correlators of $J_\Gamma$
can be computed using the  propagator $G_\chi$:
\bea
C(\vec{p},t) &=& \left\langle\sum_{\vec{x}} e^{-i\vec{p}\cdot\vec{x}}
\;\tr_{\rm c}\Big\{G_\chi(\vec{x},t;\vec{0},0) \right. \nonumber \\
 &\times &  \tr_{\rm s}\left[ \Gamma \, \Omega(\vec{x},t) \;
 \Gamma \, G_Q(\vec{0},0;\vec{x},t)\right]\Big\} \bigg\rangle \, .
\eea

A forthcoming paper will show that heavy-light mesons in
this formulation do not have the complicated flavor structure
that light-light mesons do.
Mixing between pion operators is due to the flavor-changing
exchange of high momentum gluons;
since the heavy quark action does not have the doubling symmetry
and describes only one flavor, absorption of a gluon with momentum
component $p_i =\pi/a$
would increase the heavy quark energy by $\pi^2/(2Ma^2)$,
a large factor.
As a result, flavor-changing contributions to heavy-light
Greens functions are suppressed.

\section{PRELIMINARY RESULTS}

An exploratory calculation was performed on an
$8^3\times 20$ lattice.  200 gauge configurations were generated
using the tree-level, tadpole-improved L\"uscher-Weisz action 
within the quenched approximation.  The coupling 
$6/g_0^2 = 1.719$ corresponds to $1/a = 0.8$ GeV.
The heavy quark propagators were evolved using the
NRQCD action of Ref.\ \cite{Lepage:1992tx} with bare mass
$aM_0$ = 6.5 which roughly corresponds to the bottom quark
mass.  The light quark propagators were computed with
several bare masses, but we focus on $am_0 = 0.18$ 
which roughly corresponds to the strange sector.
We use only local sources and sinks in this work.

Since the NRQCD propagator requires only an initial condition,
the heavy-light correlators are not periodic in $t$, and
amplitudes and energies can only be extracted from $C(\vec{p},t)$ 
by fitting the data with $t\in[0,9]$.  Since the 
opposite parity state contributes
to the correlator with a factor of $(-1)^{t+1}$, we must
include at least two terms in the fit.  In fact, a third is
necessary since it appears that the first excited state is
lower in energy than the parity partner.  Thus it is desirable
to utilize the constrained curve fitting discussed
in Ref.\ \cite{Lepage_lat01}.

We extract energies and amplitudes by fitting the data to
the form
\bea
C(\vec{p},t)&&= {A_0} e^{-{E_0} t} 
~+~ (-1)^{t+1} {A_1} e^{-{E_1} t} \nonumber \\
+ \sum_{n=2}^{N}&& (-1)^{n(t+1)} {A_n} 
e^{-(\Delta E_n + \Delta E_{n-2} + \ldots) \,t}
\label{eq:fitform}
\eea
where $\Delta E_n \equiv E_n - E_{n-2}$.
Gaussian priors were used for the fit parameters, with
widths as follows: $\delta A_{0,1} = 50\%$, 
$\delta E_{0,1} = 25\%$, $\delta A_{n\ge2} = 100\%$,
and $\delta (\Delta E_{n\ge2}) = 75\%$.  The prior means
for the ground state energies and amplitudes were selected
based on effective mass plots, excited state amplitude 
means were equal to or less than the ground state amplitudes,
and a rough guess was made for the excited state splittings
with the assumption that $\Delta E_n$ is roughly independent of $n$.
We use all timeslices from 0 to 9 in the fit.
Most of our fits were done with $N=3$, however fits with 
$N=4-8$ gave similar results but with
increased uncertainties (for the priors given).
When possible, we also performed a standard unconstrained fit
and found good agreement.

\begin{figure}[t]
\vspace{4.9cm}
\includegraphics{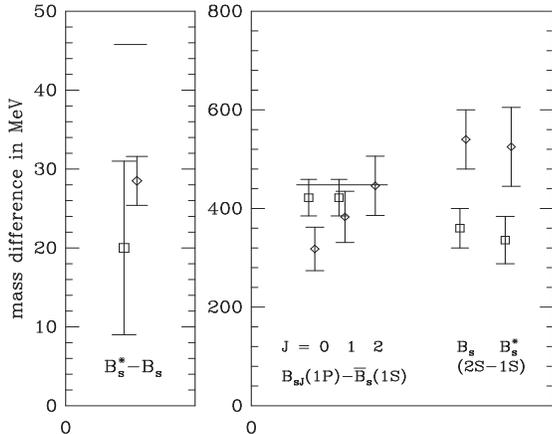}
\caption{
\label{fig:splittings}
Mass splittings in the $B$ spectrum, from either the $0^-$
or the spin averaged 1S states.
Squares mark our preliminary results, and
diamonds are taken from Ref.\ \cite{Hein:2000qu}. 
The horizontal lines correspond to experimental
observations of the $B_s^*-B_s$ splitting and
the spin-averaged $B_s(1P)-B_s(1S)$ splitting.
}
\end{figure}

Preliminary results are summarized in Fig.\ \ref{fig:splittings}.
The $1S$ and $1P$ masses agree with
 state-of-the-art NRQCD/improved-Wilson calculations 
\cite{Hein:2000qu,Lewis:2000sv} which used  finer lattice spacings.
Note that the $J=0$ and $1$ $P$-wave masses are automatically 
extracted as the parity partner states ($E_1$ in (\ref{eq:fitform}))
of the $B_s$ and $B_s^*$, respectively.
The $2S$ masses we extract from the $\Delta E_2$ in 
(\ref{eq:fitform}) are lower than from Ref.\ \cite{Hein:2000qu}.

\section{IMPROVED ACTIONS}

Significant progress has been made in the past few years
in improving the staggered fermion action, suppressing
the large $O(a^2)$ flavor-changing interactions mediated
by large momentum gluons 
\cite{Blum:1997uf,Orginos:1999ue,Lepage:1999vj}.  We
studied several staggered actions: the original 1-link
action, the Naik action \cite{Naik:1989bn}, 
the full $O(a^2)$
improved action \cite{Lepage:1999vj}, and the 
``flavor-improved'' action which is the full action
minus the Naik term.

The improvement of the pion dispersion relation can be
measured through the speed-of-light parameter
$c(|\vec{p}|^2)$:
\beq
c^2(|\vec{p}|^2) = (E^2(\vec{p}) - E^2(0))/|\vec{p}|^2 \, .
\eeq
For $a\vec{p} = (2\pi/8) \vec{k}$ (all permutations of $\vec{k}$)
we found the following preliminary values for $c^2(|\vec{p}|^2)$:
\\[.5cm]
\centerline{
\begin{tabular}{l|cc}
action & $\vec{k} = (0,0,1)$ & $\vec{k} = (0,1,1)$  \\ \hline
1-link & 0.651(14) & 0.629(11) \\
Naik & 0.882(14) & 0.908(36) \\
flavor & 0.721(48) & 0.710(37) \\
$O(a^2)$ & 0.894(26) & 0.924(24)
\end{tabular}
}
\\[.5cm]
The Naik term is clearly responsible for improving the
pion dispersion relation.

A good fit to the heavy-light 
correlators could not be obtained with the Naik and $O(a^2)$
actions.  We believe this is due to the unphysical negative
norm states present as a consequence of the 3-link coupling
in the $\hat{t}$ direction.  This hypothesis is supported by
two further observations: the $\chi^2$ of the fits decreased
when the priors allowed for some terms to have negative amplitudes,
and good fits were obtained using a fifth action which was 
fully  $O(a^2)$ improved in the spatial directions, but had no
Naik term in the $\hat{t}$ direction.  It is an open question
as to why the temporal Naik term affects heavy-light correlators
much more than light-light correlators.

\section{OUTLOOK}

We have demonstrated the feasibility of using staggered 
fermions as the light quark action in a heavy-light system.
The Bayesian method of constrained curve fitting is a useful,
and in a few instances, necessary tool for fitting correlators
on 10 time-slices to 3 or more exponentials.
The spectrum on a coarse lattice agrees for the most part with
other existing results, although the $B_s(2S)$ mass extracted
from our fits does not.  Calculation of the non-strange $B$ 
spectrum is in progress.
Once the appropriate renormalization constants are computed, 
this method will be useful for computing decay constants and 
semi-leptonic form factors.

\section*{ACKNOWLEDGMENTS}

This work was supported in part by the U.S.\ Department of Energy.
We are grateful to the MILC collaboration for their
public code \cite{MILCcode}
which was the basis for the calculation of
staggered fermion propagators, and to the Center for
Computational Physics at the University of Tsukuba,
where part of this work was done.  Numerical simulations
were performed at the Ohio Supercomputer Center.



\begin{thebibliography}{15}

\bibitem{Lepage:1992tx}
G.~P.~Lepage, L.~Magnea, C.~Nakhleh, U.~Magnea and K.~Hornbostel,
Phys.\ Rev.\ D {\bf 46}, 4052 (1992).

\bibitem{Lepage_lat01}
G.P.\ Lepage, plenary talk at this conference.

\bibitem{Hein:2000qu}
J.~Hein {\it et al.},
Phys.\ Rev.\ D {\bf 62}, 074503 (2000).

\bibitem{Lewis:2000sv}
R.~Lewis and R.~M.~Woloshyn,
Phys.\ Rev.\ D {\bf 62}, 114507 (2000).

\bibitem{Blum:1997uf}
T.~Blum {\it et al.},
Phys.\ Rev.\ D {\bf 55}, 1133 (1997).

\bibitem{Orginos:1999ue}
K.~Orginos and D.~Toussaint,
Phys.\ Rev.\ D {\bf 59}, 014501 (1999).

\bibitem{Lepage:1999vj}
G.~P.~Lepage,
Phys.\ Rev.\ D {\bf 59}, 074502 (1999).

\bibitem{Naik:1989bn}
S.~Naik,
Nucl.\ Phys.\ B {\bf 316}, 238 (1989).

\bibitem{MILCcode}
http://physics.indiana.edu/$\sim$sg/milc.html\,.

\end{thebibliography}
\end{document}